\documentclass[11pt, a4paper]{article}
\usepackage{graphicx}
\usepackage{amssymb}
\usepackage{amsmath}
\usepackage{color}
\usepackage{setspace}
\usepackage{marvosym}
\usepackage{upgreek}
\usepackage{braket}
\usepackage{nicefrac}
\usepackage{textcomp}
\usepackage[margin=0.8in]{geometry}
\usepackage{booktabs}
\usepackage{multirow}
\usepackage[section]{placeins} 
\usepackage{lineno}
\usepackage{multicol}
\usepackage{caption}
\newenvironment{Figure}
  {\par\medskip\noindent\minipage{\linewidth}}
  {\endminipage\par\medskip}
\begin{document}  %\linenumbers

\begin{center}
\textbf{\Large{High-fidelity entanglement between a trapped ion and a telecom photon via quantum frequency conversion}}\\
\vspace{10pt}
Matthias Bock\footnote{These authors contributed equally to this work \label{fnt:note1}}, Pascal Eich$^{\ref{fnt:note1}}$, Stephan Kucera, Matthias Kreis, Andreas Lenhard, \\ Christoph Becher\footnote{christoph.becher@physik.uni-saarland.de \label{fnt:note2}}, and J\"{u}rgen Eschner\footnote{juergen.eschner@physik.uni-saarland.de \label{fnt:note3}} \\ \vspace{1em}
\textit{Universit\"at des Saarlandes, Fachrichtung Physik,  Campus E2.6, 66123 Saarbr\"ucken, Germany\\
} \end{center} \vspace{1em}
\small
\setlength{\baselineskip}{12pt}
\begin{multicols}{2}
\textbf{Entanglement between a stationary quantum system and a flying qubit is an essential ingredient of a quantum-repeater network. It has been demonstrated for trapped ions, trapped atoms, color centers in diamond, or quantum dots. These systems have transition wavelengths in the blue, red or near-infrared spectral regions, whereas long-range fiber-communication requires wavelengths in the low-loss, low-dispersion telecom regime. A proven tool to interconnect flying qubits at visible/NIR wavelengths to the telecom bands is quantum frequency conversion. Here we use an efficient polarization-preserving frequency converter connecting 854$\,$nm to the telecom O-band at 1310$\,$nm to demonstrate entanglement between a trapped $^{40}$Ca$^{+}$ ion and the polarization state of a telecom photon with a high fidelity of 98.2 $\pm$ 0.2$\%$. The unique combination of 99.75 $\pm$ 0.18\% process fidelity in the polarization-state conversion, 26.5$\%$ external frequency conversion efficiency and only 11.4 photons/s conversion-induced unconditional background makes the converter a powerful ion-telecom quantum interface.}\\

\begin{center}\noindent\textbf{Introduction}\normalsize\\ \end{center}

Quantum repeaters that establish long-distance entanglement are essential tools in the emerging field of quantum communication technologies \cite{cizoll}. While proposals for memoryless repeaters exist (e.g.\ \cite{munro}), many currently pursued approaches require efficient, low-noise quantum memories as nodes that exchange quantum information via photonic channels \cite{sang}. Various atomic and solid-state systems have been identified as suitable quantum nodes, e.g. trapped ions \cite{blinov, stute}, trapped atoms \cite{volz, ritter}, color centers in diamond \cite{togan} or quantum dots \cite{degreve, gao}. Their optical transitions, however, are --with few exceptions-- located outside the wavelength regime between 1260$\,$nm and 1625$\,$nm, where telecom fibres afford low-loss transmission. Thus, there is a demand for interfaces connecting the telecom-wavelength regime and the visible/NIR range in a coherent way, i.e., preserving quantum information encoded in a degree of freedom of a single photon, such as its polarization.\\
Promising candidates for such interfaces are, e.g., non-degenerate photon-pair sources \cite{bussieres, seri, lenhard} or quantum frequency converters (QFC) \cite{kumar}. The latter can be implemented either by four-wave mixing (FWM) using resonances in cold atomic ensembles \cite{radnaev, dudin} or by a solid-state approach utilizing three-wave mixing in $\chi^{2}$- or four-wave mixing in $\chi^{3}$-nonlinear media \cite{srinivasan}. It has been shown that $\chi^{2}$-based QFC preserves nonclassical properties of single photons and photon pairs, such as second-order coherence \cite{zaske, ates, rakher}, time-energy entanglement \cite{tanzilli}, time-bin entanglement \cite{ikuta}, orbital angular momentum entanglement \cite{guooam}, polarization entanglement \cite{ramelow}, and photon indistinguishability \cite{kambs, langrock}; furthermore, nonclassical correlations between telecom photons and spin waves in cold atomic ensembles \cite{farrera, ikuta2} have been demonstrated. Using near-resonant QFC based on FWM in an atomic ensemble, entanglement of a spin wave qubit with the polarization state of a telecom photon has been realized \cite{dudin}. A corresponding implementation using solid-state QFC has remained an open challenge, despite being a highly desirable approach for its wavelength flexibility: while atomic ensembles are restricted to the particular transition wavelengths of neutral atoms, solid-state QFC can be adjusted to the system wavelength of other promising stationary quantum bits for quantum nodes, such as trapped ions, color centers in diamond or rare-earth ensembles. The main obstacle has been the strong polarization dependence of the $\chi^{2}$-process and the high demands on efficiency and noise properties of the converter. Despite successful attempts to overcome the polarization dependency \cite{ramelow, albota, krut}, the integration of a solid-state QFC device that fulfills all above mentioned requirements into a quantum node has not been achieved.\\
%%%%%%%%%%%%%%%%%%%%%%%%%%%%%%%%%%% 
\begin{figure*}[tb]
	\centering
	\includegraphics[width=1.0\textwidth]{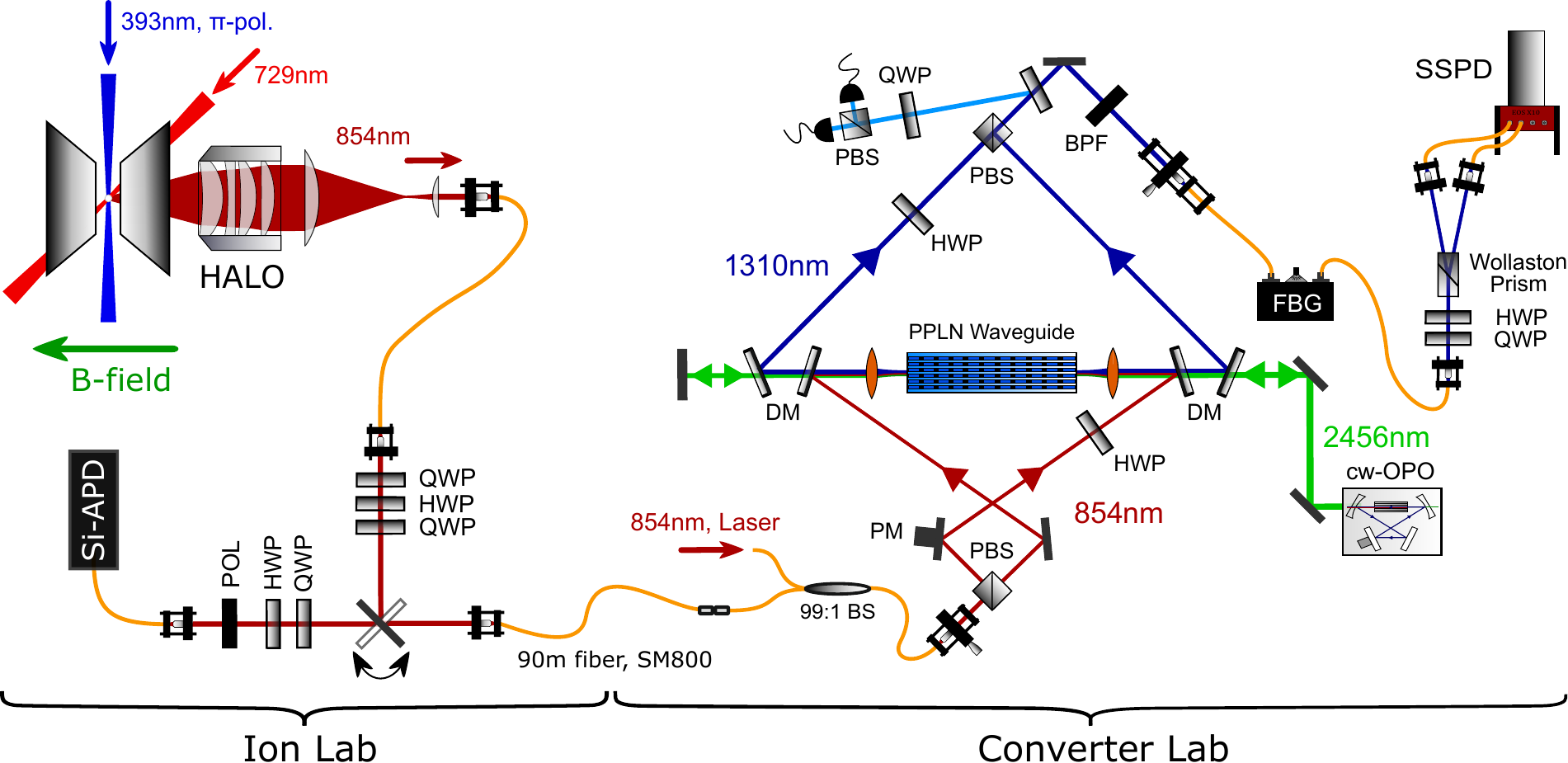}
	\caption{\textbf{Experimental setup.} Atom-photon entanglement is generated between a single trapped $^{40}$Ca$^{+}$-ion, confined and laser-cooled in a linear Paul trap, and a single photon at 854$\,$nm. The photons are collected with a HALO ("High numerical-Aperture Laser Objective") and coupled to a single mode fiber. A combination of two quarter wave plates (QWP) and one half wave plate (HWP) is inserted behind the fiber to compensate all unitary rotations of the polarization state caused by the single mode fiber and several dichroic mirrors between the fiber and the HALO. A flip mirror can be inserted behind the fiber to send the photons either to the converter lab via a 90$\,$m long single mode fiber or to the projection setup for 854$\,$nm. The polarization-preserving frequency converter is realized with a nonlinear waveguide crystal in a single-crystal Mach-Zehnder configuration (for details see main text and Methods). The converted photons are detected with superconducting single-photon detectors. Further abbreviations: PBS: polarizing beam splitter, DM: dichroic mirror, BS: beam splitter, BPF: band pass filter FBG: fiber Bragg grating.}
	\label{fig:setup}
\end{figure*}
%%%%%%%%%%%%%%%%%%%%%%%%%%%%%%%%%%% 
Single trapped ions are promising systems for quantum nodes, providing a very high level of control over their photonic interaction \cite{stute, kurz1, kurz2} and large coherence times \cite{ruster, wang}; importantly, single-ion qubits are directly addressable and thus allow quantum information processing via high-fidelity quantum gates \cite{ballance, gaebler}. In our work we connect a trapped-ion quantum node via QFC to the telecom regime in a coherent way, creating high-quality entanglement between the ion and a telecom photon. To this end, we generate entanglement between an atomic quantum bit in a single trapped $^{40}$Ca$^{+}$ ion and the polarization state of a single photon at 854$\,$nm. Subsequent polarization-preserving QFC to the telecom O-Band establishes high-fidelity entanglement between ion and telecom photon, which we verify by quantum state tomography.\\

\begin{center}\noindent\textbf{Results}\normalsize\\ \end{center}

\noindent \textbf{Ion-photon quantum interface.} The ion-photon interface is shown in Fig.~1. A single $^{40}$Ca$^{+}$ ion is confined in a linear Paul trap and laser-excited. Photons at 854\,nm emitted along the quantization axis --defined by a magnetic field-- are collected with 3.6\% efficiency by a HALO ("High numerical-Aperture Laser Objective", NA = 0.4) and coupled to a single-mode fiber with 39\% efficiency (for details see methods section). Further details on the setup are found in earlier publications \cite{kurz1, kurz2, schug}. The experimental sequence to generate atom-photon entanglement is shown in Fig.~2a. Starting from the ground state $S_{1/2}$, the ion is excited to the short-lived $P_{3/2}$ state with $\pi$-polarized laser light at 393$\,$nm. Spontaneous decay to $D_{5/2}$ leads to entanglement between the atomic states $\Ket{-\nicefrac{3}{2}}$ = $\Ket{D_{5/2}, m=-\nicefrac{3}{2}}$ and $\Ket{\nicefrac{1}{2}}$ = $\Ket{D_{5/2}, m=\nicefrac{1}{2}}$ and the emitted 854$\,$nm photon in the polarization states $\Ket{R}$ and $\Ket{L}$. For details on the sequence see Supplementary Note 1. Taking into account the Clebsch-Gordan coefficients (CGC) of the two transitions (see Fig.~2a), the ideal ion-photon state is
\begin{equation}
	\Ket{\mathit{\Psi_{\text{ideal}}}} = \sqrt{\frac{2}{3}}\Ket{R, -\nicefrac{3}{2}} + \sqrt{\frac{1}{3}}\Ket{L, \nicefrac{1}{2}}~.
\end{equation}
The experimentally generated ion-photon state is characterized by quantum-state tomo-graphy (see methods). The real and imaginary parts of the reconstructed density matrix $\rho$ are shown in Fig.~2b. From $\rho$ we deduce the fidelity $F = \Bra{\mathit{\Psi_{\text{ideal}}}}\rho\Ket{\mathit{\Psi_{\text{ideal}}}}$, denoting the overlap between the generated and the ideal state, and the purity $P = {\rm Tr}(\rho^{2})$, a measure for the depolarization of the state. We find $F = 98.3 \pm 0.3\%$ and $P = 96.7 \pm 1.6\%$. An upper bound of the fidelity for a given purity is $F_{\text{max}}$ = $\frac{1}{2}$(1+$\sqrt{2P-1}$) = 98.3\% indicating that the fidelity is solely limited by depolarization and not by undesired unitary rotations of the state \cite{bussieres}. Depolarization is mainly caused by polarization-dependent loss in the optics behind the ion trap; minor contributions arise from non-perfect readout pulses and loss of atomic coherence. Calculating the overlap with a maximally entangled Bell state yields $F_{\text{Bell}} = 95.5 \pm 0.3\%$. The maximum possible value of $F_{\text{Bell}}$ for our state is 97\% (for $P = 1$), due to the asymmetric CGCs. Note that these numbers are calculated after subtraction of the detector dark counts, in order to characterize the functionality of our method. In a realistic repeater scenario these dark counts have to be included. Without background subtraction we deduce a purity $P = 92.1 \pm 1.6 \%$ and fidelities $F = 95.9 \pm 0.3 \%$ and $F_{\text{Bell}} = 93.3 \pm 0.3 \%$. Even without background subtraction, all fidelities are many standard deviations above the classical threshold of 50\%, as well as above the threshold of 70.7\% necessary to violate Bell's inequalities. The method to perform the background subtraction along with tables summarizing all fidelities and purities can be found in Supplementary Note 7 and Supplementary Tables 1,2 \& 3.\\  
\vspace{1.5em}
%%%%%%%%%%%%%%%%%%%%%%%%%%%%%%%%%%%%%
\begin{Figure}
	\centering
	\includegraphics[width=0.9\textwidth]{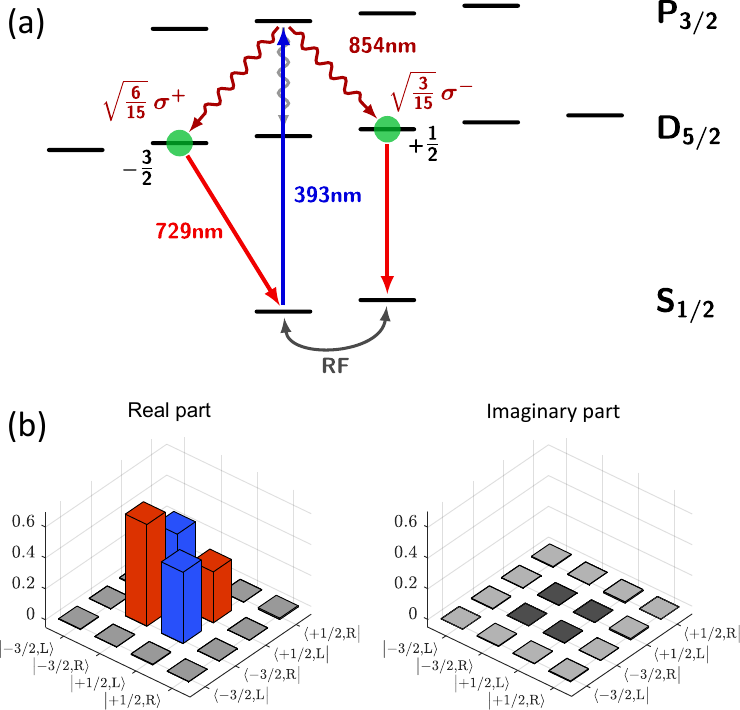}
	\captionof{figure}{\textbf{Ion-photon entanglement scheme and quantum state tomography.} (a)~ Atom-photon-entanglement is generated via spontaneous decay from the $P_{3/2}$- to the $D_{5/2}$-state after excitation with $\pi$-polarized laser light at 393$\,$nm. Emitted photons at 854$\,$nm are collected along the quantization axis, thereby suppressing $\pi$-polarized photons. Atomic state analysis is realized via coherent pulses on the optical transition at 729$\,$nm and the RF-transition in the ground state followed by fluorescence detection (details see Supplementary Note 1). (b) Real and imaginary part of the density matrix of the ion-photon entangled state, measured via quantum state tomography. The different heights of the diagonal elements (red bars) results from the different Clebsch-Gordan coefficients of the $\sigma^{+}$- and the $\sigma^{-}$-transitions.\\}
	\label{fig:resno}
\end{Figure}
%%%%%%%%%%%%%%%%%%%%%%%%%%%%%%%%%%%%%
%\vspace{1em}
With a sequence repetition rate of $\sim 58$\,kHz, we obtain 236 generated (27.6 projected and detected) entanglement events/s, which compares well with other Ca$^{+}$-ion systems \cite{stute}. One order of magnitude higher entanglement rates were reported in a Yb$^{+}$-system, mainly due to a higher sequence repetition rate enabled by shorter cooling times and the use of ultrafast laser pulses \cite{hucul}. Our signal-to-background ratio (SBR) is 29.5, solely limited by detector dark counts. A detailed account of the derivation of all numbers is given in Supplementary Notes 4 \& 5.\\

\noindent \textbf{Polarization-preserving quantum frequency converter.} The ion-trap setup is connected by 90\,m of fiber to the converter setup, where the ion-entangled photons at 854\,nm are converted to 1310\,nm employing a periodically-poled lithium niobate (PPLN) ridge waveguide designed for the DFG-process $\nicefrac{1}{\text{854$\,$nm}}$ - $\nicefrac{1}{\text{2456$\,$nm}}$ = $\nicefrac{1}{\text{1310$\,$nm}}$. As shown in Fig.~1, the polarization selectivity of the DFG-process is overcome using a polarization interferometer: an arbitrary input state is split into H- and V-polarizations, a HWP rotates the H-polarization to the convertible V-polarization, and both are coupled via dichroic mirrors and zinc selenide aspheric lenses to the same wave\-guide from opposite directions. The V-polarized strong pump field at 2456$\,$nm, generated by a home-built optical parametric oscillator (OPO, see methods) is aligned in double-pass configuration to facilitate conversion in both directions. The two converted polarizations are separated with dichroic mirrors from the other beams and superimposed on a second PBS after undoing the rotation from H to V with another HWP. To ensure faithful conversion of arbitrary input polarizations, the path length of the interferometer is actively stabilized by injecting light from a stabilized diode laser at 854\,nm via a chopper wheel and a 99:1 beam splitter into the setup. A second 99:1 BS splits a part of the converted laser light from the photon path, with which we measure the path length difference (light blue beam path). Feedback on the path length is realized by a piezo actuator connected to one of the mirrors (PM). Spectral filtering of the converted light with a broadband bandpass filter (25$\,$nm) and a narrowband fiber Bragg grating (25$\,$GHz) suppresses the remaining pump light as well as noise arising from non-phase-matched nonlinear processes. With these two filter stages, the conversion-induced unconditional noise is reduced to 11.4 photons/s. The external conversion efficiency $\eta_{\text{ext}}$ (defined as "fiber-to-fiber" efficiency of the complete QFC device) of the two interferometer arms vs.\ the power of the pump field at 2456\,nm, $P_{\text{P}}$, is shown in Fig.~3a. The data points are described quite well by the theoretical curve $\eta_{\text{ext}}\left(P_{\text{P}}\right) = \eta_{\text{ext,max}} \text{sin}^{2}\left(\sqrt{\eta_{\text{nor}}P_{\text{P}}}L\right)$ \cite{zaske2}. However, the setup is not fully symmetric with respect to forward and backward conversion: due to losses in the optics behind the waveguide, the backwards-propagating pump power is lower, thus the curve of the V-polarized arm (red data points) is shifted to higher pump powers. Nevertheless, we identify a working point at the intersection of the curves, which ensures an equal conversion efficiency of 26.5\% for H- and V-polarized light, which compares well with other QFC systems \cite{zaske, rakher, ikuta, kambs, krut}. In order to verify that the converter preserves arbitrary input polarization states, we apply process tomography \cite{chuang} using laser photons. We prepare four different input states \{H,V,D,L\} and measure the respective Stokes vectors of the output state. With that the process matrix $\chi$ in the Pauli basis, connecting the in- and output density matrices via $\rho_{\text{out}}$ = $\sum_{\text{mn}} \chi_{\text{mn}}\sigma_{\text{m}}\rho_{\text{in}}\sigma_{\text{n}}^{\dagger}$, is calculated (Fig.~3b). In the ideal case, $\chi$ possesses only a single non-zero entry $\chi_{00}$ denoting the identity operation. This entry can be identified as process fidelity, which in our case is $F_{\text{pro}}$ = 99.75 $\pm$ 0.18\%, confirming very high-fidelity conversion of the input polarization state. The error in $F_{\text{pro}}$ is deduced from a Poissonian distribution and arises from power fluctuations of the input and the pump field. Further details on the converter are given in the method section.\\
%%%%%%%%%%%%%%%%%%%%%%%%%%%%%%%%%%%%%
\vspace{1.5em}
\begin{Figure}
	\centering
	\includegraphics[width=0.6\textwidth]{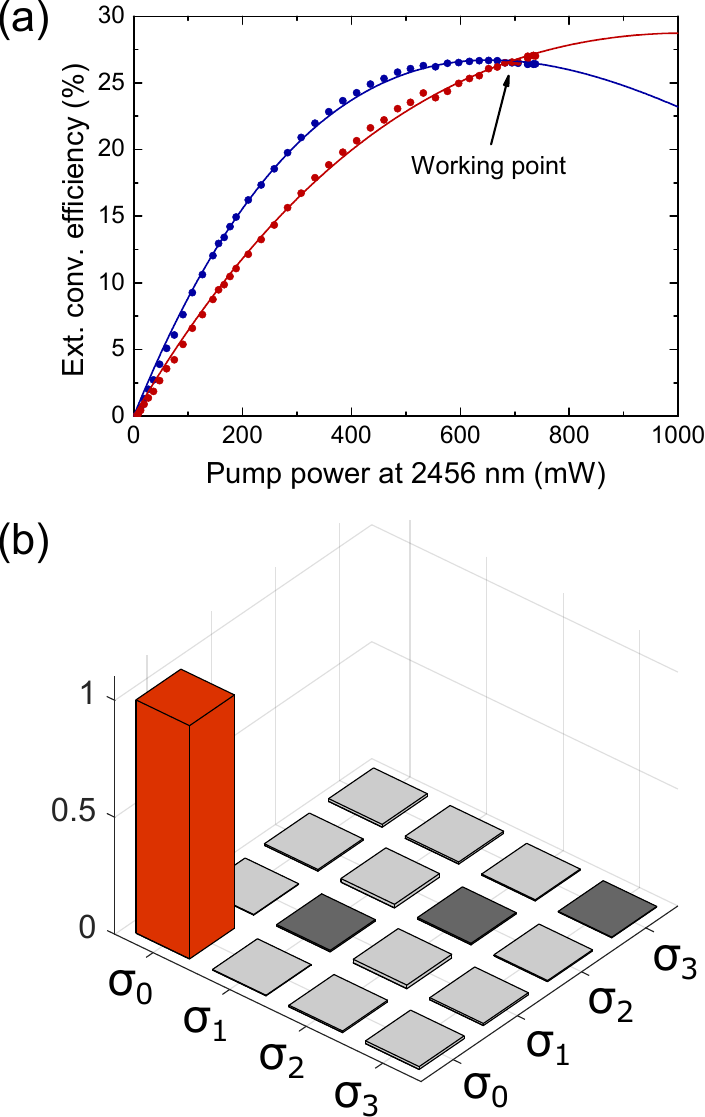}
	\captionof{figure}{\textbf{Characterization of the quantum frequency converter.} (a)~The external conversion efficiency of the two interferometer arms depending on the pump power of the mixing field at 2456 nm. The data points are fitted with $\eta_{\text{ext}}\left(P_{\text{P}}\right) = \eta_{\text{max}} \text{sin}^{2}\left(\sqrt{\eta_{\text{nor}}P_{\text{P}}}L\right)$ \cite{zaske2}. (b) The absolute values of the process matrix of the coherent polarization-preserving down-conversion measured with a laser. The process fidelity is determined by element corresponding to the identity operation (red bar). We achieve a value of 99.75$\pm$0.18\%. The error bars are deduced from a Poissonian distribution.\\}
	\label{fig:char}
\end{Figure}
%%%%%%%%%%%%%%%%%%%%%%%%%%%%%%%%%%%%%
\noindent \textbf{Ion-telecom-photon entanglement.} To characterize the full quantum interface we investigate the performance of the combined ion-converter system: detecting the telecom output photons on a superconducting single-photon detector (SSPD) yields 43.5 generated (24.8 projected/detected) events/s with a SBR of 24.3. These numbers are in very good agreement with the previously determined conversion and detection efficiencies (see Supplementary Notes 4 \& 5). Despite the loss in the conversion, the SBR is only weakly affected as we benefit from the detector's higher efficiency and lower dark-count rate. The density matrix of the ion-photon state after conversion is depicted in Fig.~4a, yielding $F = 97.7 \pm 0.2$\%, $P = 95.8 \pm 1.3$\%, and $F_{\text{Bell}}$ = 94.8 $\pm$ 0.2\% after background subtraction. This result unambiguously verifies the entanglement between the ion and the telecom photon after QFC. The reduction of the fidelity by 0.6\% compared to the unconverted ion-photon state is higher than what we expect from the process fidelity. We attribute this to power fluctuations of the pump laser and slow polarization drifts in the fiber connecting the setups. The background in these measurements has two contributions: a minor part of 6.5\% due to conversion-induced noise and a major part of 93.5\% stemming from detector dark counts (Supplementary Note 7). To quantify the influence of the converter on the final entangled state, it is useful to consider the case when only the detector part of the background is subtracted: we obtain $P = 95.1 \pm 1.3 \%$, $F = 97.3 \pm 0.2 \%$ and $F_{\text{Bell}} = 94.5 \pm 0.2 \%$, which confirms that the conversion-induced noise has only a minor influence on the final state. If no background subtraction is applied, we get $P = 90.3 \pm 1.2 \%$, $F = 94.8 \pm 0.2 \%$ and $F_{\text{Bell}} = 92.2 \pm 0.2 \%$.\\ 
\vspace{1.5em}
%%%%%%%%%%%%%%%%%%%%%%%%%%%%%%%%%%%%%%%%%%%%%
\begin{Figure}
	\centering
	\includegraphics[width=0.9\textwidth]{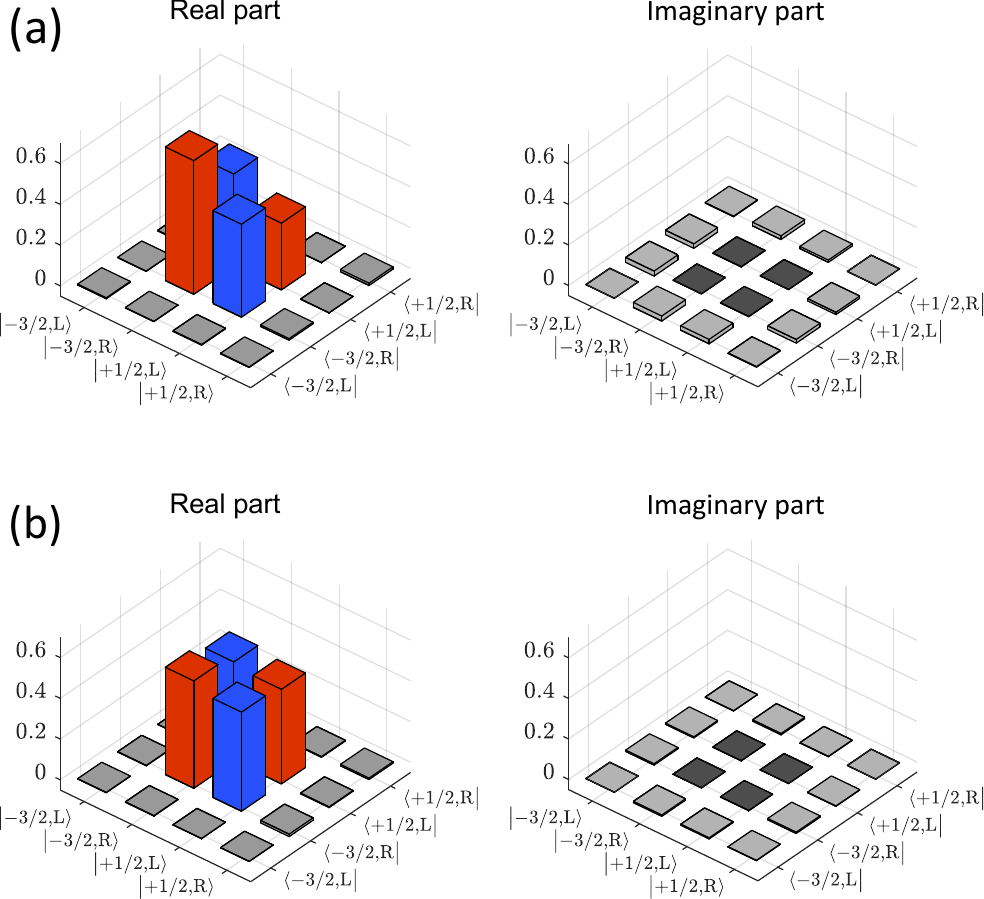}
	\captionof{figure}{\textbf{Quantum state tomography of the ion-telecom-photon entangled states.} Real and imaginary parts of the density matrices for: (a) the converted "bare" ion-photon entangled state still revealing the asymmetry in the diagonal elements (red bars) due to the asymmetric Clebsch-Gordan coefficients (b) the converted ion-photon entangled state projected onto a Bell state by introducing polarization-dependent losses.\\}
	\label{fig:rescon}
\end{Figure}
%%%%%%%%%%%%%%%%%%%%%%%%%%%%%%%%%%%%%%%%%%%%%
%\vspace{0.5em}
Beyond the faithful QFC of ion-photon entanglement, the converter also renders possible the generation of maximally entangled states: we realize this by rotating the polarization of the 854$\,$nm photons in a way that $\Ket{R}$ and $\Ket{L}$ correspond to the converter's interferometer arms. Then we reduce the conversion efficiency of the $\Ket{R}$ arm by a factor of two to compensate the higher CGC, at the cost of one third of the photons. The resulting measured density matrix is displayed in Fig.~4b. The asymmetry in the diagonal elements disappeared, and we obtain $F_{\text{Bell}} = 98.2 \pm 0.2$\%, $P = 96.7 \pm 1.4\%$ (after subtraction of only detector dark counts: $F_{\text{Bell}} = 97.7 \pm 0.2$\% and $P = 95.8 \pm 1.4\%$; without background subtraction: $F_{\text{Bell}} = 93.4 \pm 0.2$\%, $P = 87.8 \pm 1.3\%$). Thus, within the error bars, we have created a Bell state between ion and telecom photon with the same purity as the initially generated state, which proves that the converter leaves the ion-photon entanglement practically unaltered. Note that fidelity and purity in this measurement run are also in accordance with the process fidelity, which we attribute to a slightly more stable operation of the converter.\\

\begin{center}\noindent\textbf{Discussion}\normalsize\\ \end{center}

Our results demonstrate the operation of a complete quantum node that produces entangled states between a single trapped Ca$^{+}$ ion and a fiber-coupled telecom photon with a high fidelity. This constitutes a step towards the implementation of a fiber-based repeater node consisting of two ions in remote traps. In future experiments, the entanglement generation rate might be enhanced with a cavity \cite{stute}. Furthermore, conversion to the telecom C-band at 1550$\,$nm wavelength \cite{krut} shall be pursued for lower transmission losses enabling remote entanglement of ions over hundreds of kilometer fiber length. Moreover, spectral filters with narrower bandwidth combined with SSPDs with ultra-low dark-count rates \cite{tang} will lead to a much reduced background and higher SBR. Beyond these efforts, our techniques are transferable to a wide range of relevant platforms for quantum networks, such as other trapped-ion species (Yb$^{+}$, Ba$^{+}$), neutral atoms (Rb, Cs), color centers in diamond (NV$^{-}$, SiV$^{-}$), rare-earth ions in solids (Pr$^{3+}$, Nd$^{3+}$, Eu$^{3+}$), or quantum dots. Eventually, this approach opens the possibility to implement hybrid networks by coupling different quantum systems via a common bus wavelength in the telecom regime \cite{mahring}.\\

\noindent\textit{Note added:} Recently, we became aware of a related experiment by Ikuta \textit{et al.} demonstrating entanglement between a cold atomic ensemble and a telecom photon via solid-state QFC \cite{ikuta3}.\\

%\noindent \textbf{Methods:}\\

\begin{center}\noindent\textbf{Methods}\normalsize\\ \end{center}

\noindent \textbf{Photon collection from the ion.} Single 854\,nm photons emitted from the ion are collected by an in-vacuum high-numerical-aperture laser objective (HALO, Linos) with numerical aperture NA = 0.4 at a working distance of 13\,mm from the ion, covering about 4\% of the full solid angle. Collection efficiencies for photons emitted on the $\pi$ and $\sigma$ transitions are given by the dipole emission pattern. Orienting the quantization axis along the ion-HALO axis, the unnormalized polarization states of the emitted radiation are $|\psi^{(854)}_{\pi}\rangle = -\sin\theta\, |\hat{\theta}\rangle$ for $\Delta m = 0$ and $|\psi^{(854)}_{\sigma^{\pm}}\rangle = \frac{e^{\pm i \varphi}}{\sqrt{2}}\, \left( \cos\theta |\hat{\theta}\rangle \pm i |\hat{\varphi}\rangle \right)$ for $\Delta m = \pm 1$, where $\theta$ and $\varphi$ are the spherical polar and azimuthal angles of the direction of emission and $|\hat{\theta}\rangle$ and $|\hat{\varphi}\rangle$ are their respective spherical-coordinate unit vectors \cite{blinov}. Thus, the collection of photons emitted on the $\pi$ transition is suppressed due to the single-mode fiber coupling, while the collection efficiency for the $|\Delta m| = 1$ transitions sums up to 6\% with respect to spontaneous emission into full space. Taking into account the CGCs for $\sigma$- and $\pi$-decay, the resulting collection probability for $\sigma$ emission is $0.6 \cdot 6\% = 3.6\%$. The single-mode--fiber coupling accounts for an additional factor of 39\%, resulting in a total collection efficiency $\eta_{\text{coll., tot.}} \approx 1.4$\%\\

\noindent \textbf{OPO system at 2456$\,$nm.} We employ a home-built continuous-wave optical parametric oscillator (OPO) delivering 1$\,$W of single-mode, single-frequency output power at 2456$\,$nm as pump source for the DFG process \cite{lenhard2}. The OPO is pumped by a diode laser at 1081$\,$nm (Toptica DL Pro) amplified with a Yb-doped fiber amplifier (LEA Photonics) with 15$\,$W maximum output power. The OPO consists of a 40$\,$mm long periodically poled LiNbO$_{\text{3}}$ crystal with 7 poling periods ($\Lambda$ = 31.7$\upmu$m ... 32.7$\upmu$m) inside a signal-resonant bow-tie ring cavity. Tuning of the idler wavelength from 2310$\,$nm to 2870$\,$nm is achieved by changing the poling period, the crystal temperature or the cavity length via a piezo actuator. With this tuning range, we are able to cover the whole telecom O-band from 1260$\,$nm to 1360$\,$nm with the frequency converter. During the experiment, the OPO was operated at 2456$\,$nm using a poling period of $\mathit{\Lambda}$ = 32.6$\upmu$m at a temperature of 49$^\circ$C.\\

\noindent \textbf{Polarization-preserving frequency converter.} The input light is overlapped with a diagonally polarized stabilization laser at 854\,nm in a fiber beam splitter with a transmission of 99\% for the input and 1\% for the stabilization. Behind the PBS we split the orthogonal polarization components, the H-polarization is rotated to the convertible V-polarization with a HWP. Both beams are coupled to the ridge waveguide via dichroic mirrors and aspheric zinc selenide (ZnSe) lenses with focal lengths of 11$\,$mm and broadband anti-reflective (AR) coatings for all three wavelengths. The 40$\,$mm long Zn:PPLN waveguide chip (NTT Electronics) with lateral dimensions of 9 x 16$\upmu$m consists of 12 ridge waveguides with 6 different poling periods $\mathit{\Lambda}$ = 22.60$\upmu$m ... 22.85$\upmu$m (operating point: $\mathit{\Lambda}$ = 22.70$\upmu$m, $T$ = 31$^\circ$C) and AR-coatings for all wavelengths. The chip is temperature-stabilized and mounted on a 5-axis translation stage to achieve optimal mode-matching. The pump field at 2456\,nm generated by the OPO is guided free-space to the converter. The beam passes a HWP and a rutile polarizer for power control, a 1600\,nm long-pass filter used for clean-up and a telescope made of two AR-coated spherical CaF$_{\text{2}}$-lenses to achieve best possible coupling to the waveguide's fundamental spatial mode. The transmitted pump field is back-reflected by a mirror and recoupled to the waveguide to enable conversion in both directions. The converted light at 1310\,nm is separated from the pump field with further dichroic mirrors, the former H-polarized light is back-rotated with a HWP, and both arms are superimposed with another PBS. A bulk 99:1 beam splitter separates a part of the light for the path-length stabilization. Variation of the path length causes a phase change between H and V polarization, which is measured with a QWP at 22.5$^\circ$, a PBS and two photo diodes. From the two photo diode signals we calculate the contrast $\left(\frac{I_{1}-I_{2}}{I_{1}+I_{2}}\right)$, which serves as a power-independent error signal for the PID lock. The feedback on the path length is applied with a piezo actuator (PM) mounted beneath one of the mirrors. In the output arm a bandpass filter (BPF, central wavelength: 1300$\,$nm, bandwidth: 25$\,$nm, Edmund optics) is followed by a chopper wheel, blocking alternatingly the stabilization laser and the photons, and another telescope to mode-match the beam to the fiber. Fiber coupling is realized with an AR-coated aspheric lens (f=8$\,$mm, Thorlabs). As a narrowband spectral filter a fiber Bragg grating (FBG, central wavelength tunable from 1307--1317$\,$nm, linewidth: 25$\,$GHz, Advanced optical solutions GmbH) is utilized. A drawing of the complete setup and the characterization of the converter as well as a detailed analysis of the efficiencies and losses is found in Supplementary Notes 2 \& 3.\\

\noindent \textbf{Quantum state tomography.} To perform quantum state tomography, ion and photon are projected onto all 36 combinations of eigenstates of the Pauli operators $\sigma_{x}$, $\sigma_{y}$ and $\sigma_{z}$, where $\sigma_{z}$ represents the eigenbases ($\Ket{R}$/$\Ket{L}$ for the photon and $\Ket{-\nicefrac{3}{2}}$/$\Ket{\nicefrac{1}{2}}$ for the ion) and $\sigma_{x}$/$\sigma_{y}$ the superpositions of the latter. To compare the results with and without the frequency converter, we use two tomography setups for the projective measurement of the 854$\,$nm and telecom photons (see Fig.~1). The tomography setup for 854$\,$nm is inserted into the beam path via a flip mirror and consists of QWP, HWP, polarizer and a silicon APD. The APD (SPCM-AQR-14, Perkin Elmer) has a quantum efficiency of $\eta_{\text{APD}}$ = 30\% and a dark-count rate of $\gamma_{\text{DC, APD}} = 117.7$ photons/s. The tomography setup for 1310$\,$nm is realized with QWP, HWP, Wollaston prism and two commercial  fiber-coupled superconducting-nanowire single-photon detectors (SSPD, Single Quantum). The quantum efficiencies and dark counts for SSPD1 (SSPD2) are 70(2)\% (62(2)\%) and 58.7 (56.4) photons/s, respectively. All waveplates are motorized and controlled via an Ethernet link to enable remote control of the complete experiment. The atomic state is analyzed with a combination of coherent pulses on the quadrupole transition at 729$\,$nm and the RF-transition between the $S_{1/2}$-states followed by fluorescence detection (details see Supplementary Note 1). From these measurements, the density matrix is calculated via linear state reconstruction combined with a maximum-likelihood estimation (see Supplementary Note 6).\\

\noindent \textbf{Data availability}\\ 
All relevant data are available from the corresponding author on request.\\

\noindent \textbf{Acknowledgements}\\
We thank Benjamin Kambs, Philipp M\"{u}ller and Jonas Becker for helpful discussions. This work was financially supported by the German Federal Ministry of Science and Education (Bundesministerium f\"{u}r Bildung und Forschung (BMBF)) within the project Q.com.Q (Contract No. 16KIS0127).\\

\noindent \textbf{Author Contributions}\\
M.B. and P.E conducted the experiments and analyzed the data with help from S.K.. M.B. and A.L. constructed the frequency converter, P.E. and M.K. implemented the ion-photon entanglement sequence. S.K. developed a software toolbox for the state reconstruction. J.E. and C.B. conceived and supervised the project. M.B., P.E., J.E. and C.B. wrote the manuscript with input from all authors.\\

\noindent \textbf{Competing Financial Interests}\\
The Authors declare no Competing Financial or Non-Financial Interests.

\fontsize{10}{10}\selectfont
\setlength{\baselineskip}{12pt}

\end{multicols}

\clearpage

\normalsize

\newgeometry{margin=1.0in}%{left=3cm,bottom=0.1cm}
\renewcommand\refname{Supplementary References}
\renewcommand{\figurename}{Supplementary Figure}
\renewcommand{\tablename}{Supplementary Table}

\begin{center}
\textbf{\Large{Supplementary information: High-fidelity entanglement between a trapped ion and a telecom photon via quantum frequency conversion, Bock et al.}}\\
\end{center}
\vspace{3em}

%%%%%%%%%%%%%%%%%%%%%%%%%%%%%%

\Large\noindent\textbf{Supplementary Note 1\vspace{0.3em}\\ Experimental sequence for entanglement generation and atomic state analysis}\normalsize\\

The experimental sequence, depicted in Supplementary Figure~\ref{fig:sequence}, starts with 8\,\textmu s of Doppler cooling on the $S_{1/2} \rightleftarrows P_{1/2}$ transition, including repumping of population from the $D_{5/2}$ manifold back to the $S_{1/2}$ ground state, resulting in a mixture between the $\Ket{m = +\nicefrac{1}{2}}$ and $\Ket{m = -\nicefrac{1}{2}}$ Zeeman sublevels of $S_{1/2}$. As shown in Supplementary Figure~\ref{fig:sequence}a, a laser pulse at 393\,nm of 3\,\textmu s duration excites the ion to $P_{3/2}$, where the polarization and geometry of the pulse are adjusted to drive solely the atomic $\pi$ transition. Subsequent decay into the metastable $D_{5/2}$ manifold creates a mixture of entangled atom-photon states 
\begin{equation}
	\ket{\Psi_{\text{1}}} = \sqrt{\frac{2}{3}}\Ket{R, m=-\nicefrac{3}{2}} + \sqrt{\frac{1}{3}}\Ket{L, m=\nicefrac{1}{2}}
\end{equation}
and
\begin{equation}
	\ket{\Psi_{\text{2}}} = \sqrt{\frac{1}{3}}\Ket{R, m=-\nicefrac{1}{2}} + \sqrt{\frac{2}{3}}\Ket{L, m=\nicefrac{3}{2}}
\end{equation}
where the imbalance in the amplitudes originates from unequal Clebsch-Gordan coefficients for the respective decay channels. Potential decay to $D_{3/2}$ via the emission of a photon at 850\,nm is detected as false positive events and thus eliminated in a later step.

\vspace{1.5em}
\begin{figure*}[ht]
	\centering
	\includegraphics[width=1.0\textwidth]{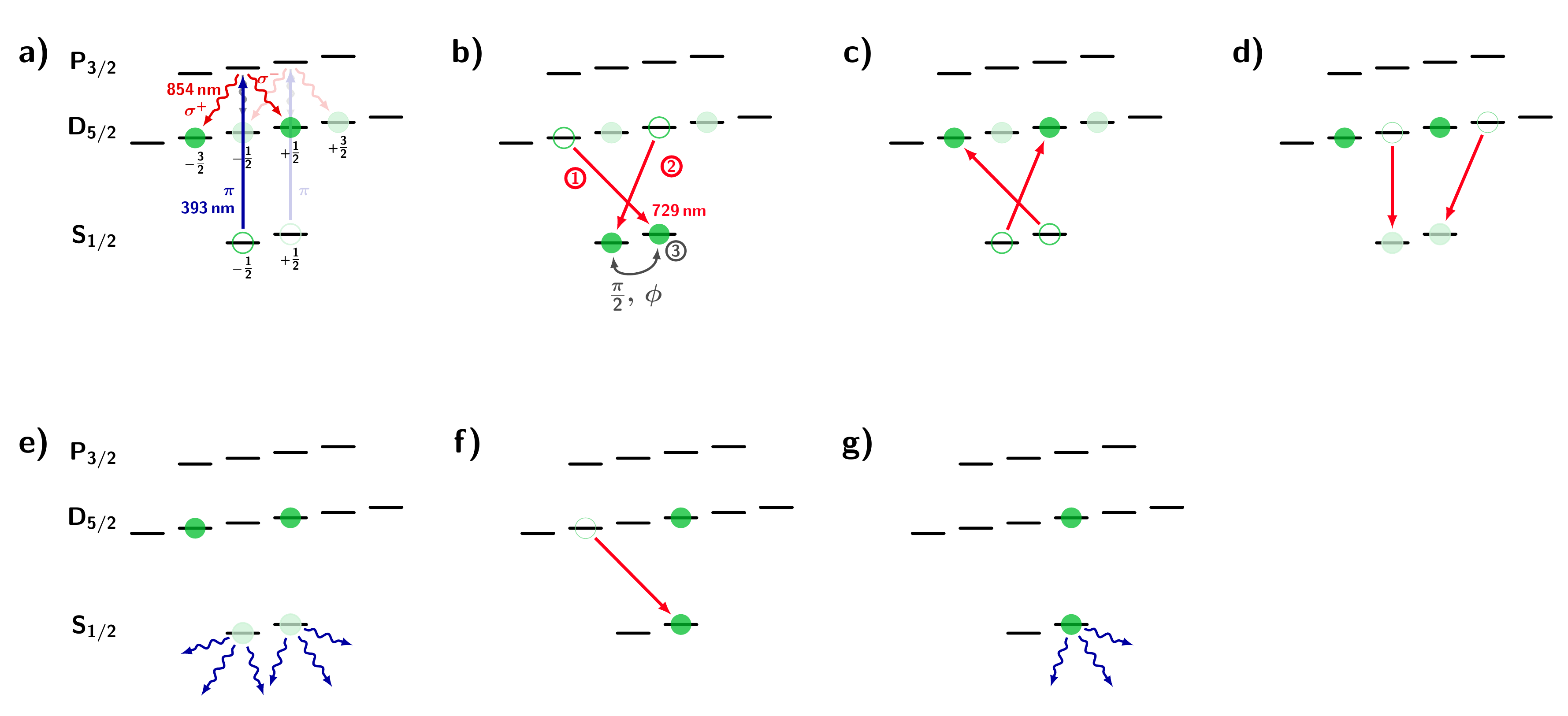}
	\caption{\textbf{Experimental sequence.} (a)~Starting from a mixture in the ground state, a mixture of two entangled atom-photon states is generated using a $\pi$-polarized laser pulse at 393\,nm. The emitted 854\,nm photon is collected along the quantization axis. (b)~Conditioned on the detection of the emitted photon, an optional RF basis-rotation pulse is applied to determine the basis for the atomic state analysis. (c)~The population is shelved back to $D_{3/2}$. (d)-(e)~Elimination of the unwanted superposition fluorescence-based state discrimination. (f)~Population in $\Ket{D_{5/2}, m=-\nicefrac{3}{2}}$ is transferred into the $S_{1/2}$ ground state. (g)~State read-out by fluorescence detection.}
	\label{fig:sequence}
\end{figure*}

As shown in Supplementary Figure~\ref{fig:sequence}b, conditioned on the detection of an 854\,nm photon, optional state rotation for the atomic state analysis is carried out by two 729\,nm pulses (50\,\textmu s each) that coherently transfer the $\Ket{m=-\nicefrac{3}{2}}$, $\Ket{m=\nicefrac{1}{2}}$ superposition into $S_{1/2}$, followed by a $\pi/2$ radio-frequency (RF) pulse (5\,\textmu s), resonant on the $\Ket{S_{1/2}, m=-\nicefrac{1}{2}} \rightarrow \Ket{S_{1/2}, m=+\nicefrac{1}{2}}$ transition, which translates the superposition phase into population of $\Ket{S_{1/2}, m=\pm\nicefrac{1}{2}}$. After shelving the population back to $D_{5/2}$ (Supplementary Figure~\ref{fig:sequence}c), the undesired superposition $\Ket{m=-\nicefrac{1}{2}}$, $\Ket{m=\nicefrac{3}{2}}$ is projected out by transferring the latter to $S_{1/2}$ and eliminating these cases by means of fluorescence detection (200\,\textmu s) where "bright" events are omitted (Supplementary Figure~\ref{fig:sequence}d/e). In the course of the fluorescence detection, erroneous 850\,nm-photon--detection events are eliminated likewise, as population in $D_{3/2}$ leads to fluorescence when the cooling lasers (397\,nm and 866\,nm) are switched on. State detection of the remaining $D_{5/2}$ populations is conducted by two further 729\,nm $\pi$ pulses coupling $\Ket{D_{5/2}, m=-\nicefrac{3}{2}}$ and $\Ket{D_{5/2}, m=\nicefrac{1}{2}}$ to $\Ket{S_{1/2}, m=\nicefrac{1}{2}}$ with a fluorescence detection after each pulse (Supplementary Figure~\ref{fig:sequence}f/g).\\
The generation sequence takes 11\,\textmu s (8\,\textmu s for cooling and 3\,\textmu s for generation), which yields a maximum achievable sequence rate of about 90.9\,kHz. However, due to the atomic state analysis, triggered upon a photon detection event, the rate is reduced to about 58\,kHz on average. 

\vspace{3em}
%%%%%%%%%%%%%%%%%%%%%%%%%%%%%%
\Large\noindent\textbf{Supplementary Note 2\vspace{0.3em}\\ Detailed setup of the quantum frequency converter}\normalsize\\

A detailed drawing of the converter, which contains all optical elements and the complete setup for the path length stabilization, is shown in Supplementary Figure~\ref{fig:setupsupp}.

\vspace{3em}
\begin{figure*}[htb]
	\centering
	\includegraphics[width=0.95\textwidth]{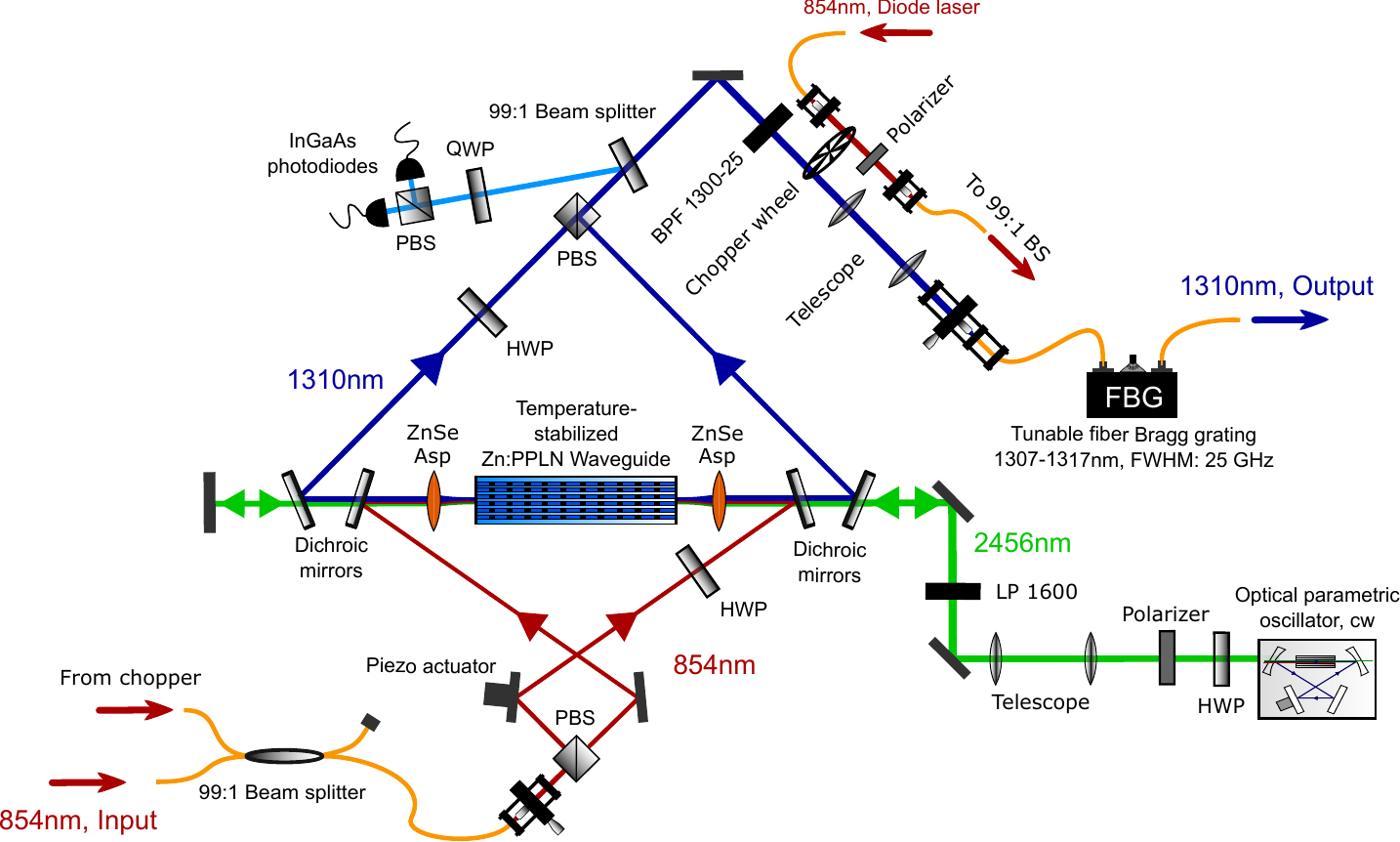}
	\caption{\textbf{Detailed drawing of the polarization-preserving frequency converter} HWP: half wave plate, QWP: quarter wave plate, PBS: Polarizing beam splitter, LP: Long-pass filter, BPF: Band-pass filter, ZnSe Asp: Zinc selenide aspheric lens}
	\label{fig:setupsupp}
\end{figure*}

\vspace{3em}        
\Large\noindent\textbf{Supplementary Note 3\vspace{0.3em}\\ External conversion efficiency}\normalsize\\

We define the external efficiency as the ratio between 1310\,nm photons leaving the output fiber of the converter (see blue output arrow in Supplementary Figure~\ref{fig:setupsupp}) and 854nm photons entering the input fiber of the converter (red input arrow in Supplementary Figure~\ref{fig:setupsupp}). Hence it includes the internal conversion efficiency and all transmission and filtering losses between these two fibers, as will be explained in more detail below. We can also interpret the external efficiency as a device efficiency of the QFC when it is applied as a component of a quantum network. Due to the asymmetry of the setup, the individual efficiencies and losses are different in the interferometer arms, which makes a separate consideration of both arms reasonable. All losses and efficiencies were measured individually using the 854\,nm alignment laser and the converted signal thereof.\\
The external conversion efficiency of the H-polarized arm (the arm containing the half wave plates) is composed of the following individual efficiencies: 
\begin{equation}
 \eta_{\text{ext$_{H}$}} = \eta_{\text{WGCoup}}\cdot \eta_{\text{InEff}}\cdot \eta_{\text{DM}}\cdot \eta_{\text{BPF}}\cdot \eta_{\text{FCoup}}\cdot \eta_{\text{FBG}}\cdot \eta_{\text{OpEle}}\cdot \eta_{\text{Asym}}  = 26.5(2)\%
\end{equation} 

$\eta_{\text{WGCoup}}$ denotes the coupling efficiency of the input field at 854\,nm to the waveguide of 79.7\%. This value is limited by the non-optimal overlap between the Gaussian mode of the input light and the astigmatic mode of the waveguide. An improvement might be achieved with cylindrical lenses, which is in our case not possible due to limited space in the 854\,nm beam path. The internal conversion efficiency $\eta_{\text{InEff}}$ is defined as the amount of input light which is converted inside the waveguide including losses in the waveguide but excluding losses outside. Not that the internal efficiency is not identical to the signal depletion, which does not take into account losses in the waveguide. We measure an internal efficiency of 96.6\% limited by the spatial mode overlap of input, pump and output field. The joint transmission of the dichroic mirrors (each beam has to pass three of them) is $\eta_{\text{DM}}$ = 81.6\%. This rather poor transmission is explained by the non-standard angles of incidence, which are outside the specification of the mirror coatings (the dichroic mirrors are specified for 45$^\circ{}$ angle of incidence). Further transmission losses arise from the spectral filters, namely the band-pass filter with $\eta_{\text{BPF}}$ = 97.1\% and the fiber Bragg grating with $\eta_{\text{FBG}}$ = 69.5\%. The coupling to the single-mode fiber $\eta_{\text{FCoup}}$ = 82\% is again limited by a non-perfect mode-matching. All losses in the remaining optical elements (lenses for fiber-coupling, ZnSe-lenses, polarizing beam splitters, 99:1 beam splitters for stabilization, spherical lenses for mode matching) are combined in $\eta_{\text{OpEle}}$ = 82.5\%. Due to a higher external efficiency of this arm, it was necessary to decrease the efficiency on purpose with the second HWP to ensure an equal conversion efficiency for both polarization components. This is represented by the asymmetry-correction factor $\eta_{\text{Asym}}$ = 92.4\%. 

We calculate the external conversion efficiency of the V-polarized arm similarly:
\begin{equation}
	 \eta_{\text{ext$_{V}$}} = \eta_{\text{WGCoup}}\cdot \eta_{\text{InEff}}\cdot \eta_{\text{DM}}\cdot \eta_{\text{BPF}}\cdot \eta_{\text{FCoup}}\cdot \eta_{\text{FBG}}\cdot \eta_{\text{OpEle}}\cdot \eta_{\text{Asym}}  = 26.6(2)\%
\end{equation} 

with the waveguide coupling efficiency $\eta_{\text{WGCoup}}$ = 78.2\%, the internal conversion efficiency $\eta_{\text{InEff}}$ = 89.6\%, the transmission through the dichroic mirrors $\eta_{\text{DM}}$ = 86.6\%, the spectral filter transmission $\eta_{\text{BPF}}$ = 97.1\% and $\eta_{\text{FBG}}$ = 69.3\%, the fiber-coupling $\eta_{\text{FCoup}}$ = 77.8\% and $\eta_{\text{OpEle}}$ = 83.8\%. Apart from small differences in the coupling efficiencies and the losses in all optical elements, a reduced internal efficiency is observed. This is caused by the lack of pump power in the second pass through the waveguide.\\

Potential improvements to boost the external conversion efficiency are waveguides with a quadratic cross-section yielding a higher overlap with Gaussian modes; a pump source with higher output powers $\geq$ 2\,W enabling pumping from both sides, which should remove the remaining asymmetries; dichroic mirrors with coatings designed for the given wavelength combination, polarization and angles of incidence which increase the transmission above 90\% and volume Bragg gratings with efficiencies around 95\% replacing the fiber Bragg grating. With these improvements, overall efficiencies around 50\% are within reach.      

%%%%%%%%%%%%%%%%%%%%%%%%%%%%%%
\vspace{3em}
\Large\noindent\textbf{Supplementary Note 4\vspace{0.3em}\\ Entanglement generation rate}\normalsize\\

\noindent\textbf{Measurement without frequency conversion}\normalsize\\

In this section, the values for the entangled-state generation rate given in the main text are calculated. In the measurement without the frequency conversion, 114,200 events were detected in 4,132\,s. Note that only the signal events excluding the APD dark counts are considered for the generation rate. The detection time window is 300\,ns corresponding to 65\% of the photon wavepacket. Thus we get for the detected entanglement rate after projection
\begin{equation}
 \gamma_{\text{854,det}} = \frac{114,200\,\text{events/s}}{4132\,s} = 27.64\,\text{events/s}
\end{equation}  
The generated rate is calculated to
\begin{equation}
 \gamma_{\text{854,gen}} = \frac{27.6\,\text{events/s}}{\eta_{\text{Pol,trans}}\cdot\eta_{\text{Pol,ab}}\cdot\eta_{\text{APD}}} = 236\,\text{events/s} 
\end{equation}  
with the transmission of the polarizer $\eta_{\text{Pol,trans}}$ = 78\%, the quantum efficiency of the APD $\eta_{\text{APD}}$ = 30\%  and an additional factor $\eta_{\text{Pol,ab}}$ = 50\% as the emitted photons are partially unpolarized, resulting in the absorption of half of the photons on average by the polarizer.\\
As a check for consistency we calculate the theoretical rate we expect from the sequence repetition rate and the losses/efficiencies between the ion trap and the projection setup. With the repetition rate $\gamma_{\text{rep}} \approx$\,58\,kHz the generated rate is  
\begin{equation}
 \gamma_{\text{854,gen,theo}} = \gamma_{\text{rep}}\cdot \eta_{\text{Halo}}\cdot \eta_{\text{Fiber}} \cdot\eta_{\text{Mix}} \cdot\eta_{\text{850}} \cdot\eta_{\text{Phot}} \approx 238\,\text{events/s}
\end{equation} 
with the collection efficiency of the HALO $\eta_{\text{Halo}}$ = 3.6\% and the fiber coupling efficiency $\eta_{\text{Fiber}} \approx$ 39\%. Furthermore three additional factors have to be taken into account. As explained in Supplementary Note 1, the state preparation creates a statistical mixture of two superpositions in the $D_{5/2}$-manifold, thus half of the photons arise from a decay to the unwanted superposition. The factor $\eta_{\text{Mix}}$ = 50\% takes into account this effect. Moreover, a decay to the $D_{3/2}$-state results in the emission of a photon at 850\,nm, which is also collected. With the Einstein coefficients of the transitions at 854\,nm and 850\,nm (A$_{\text{854}}$ = 1.35\,MHz and A$_{\text{850}}$ = 0.152\,MHz, respectively) the correction factor is $\eta_{\text{850}}$ = $\frac{\text{A}_{\text{854}}}{\text{A}_{\text{854}}+\text{A}_{\text{850}}}$ = 89.9\%. Finally, $\eta_{\text{Phot}}$ = 65\% takes into account that only a part of the photon wavepacket is considered. Thus, the theoretically expected rate is in very good agreement with the measured rate.\\ 

\vspace{2em}
\noindent\textbf{Measurement with frequency conversion}\normalsize\\

The detected entanglement generation rate including the frequency converter is computed in an analogue way. Note that we utilized two single-photon detectors for the telecom photon detection instead of a single APD for the unconverted photon detection. With 193,120 events detected in 7,779\,s we get    
\begin{equation}
 \gamma_{\text{1310,det}} = \frac{193,120\,\text{events/s}}{7,779\,s} = 24.8\,\text{events/s}
\end{equation} 
The generated entanglement rate for the converted photons is 
\begin{equation}
 \gamma_{\text{1310,gen}} = \frac{12.4\,\text{events/s}}{\eta_{\text{Tomo}}\cdot\frac{\eta_{\text{SSPD1}}+\eta_{\text{SSPD2}}}{\text{2}}} = 43.5(14)\,\text{events/s} 
\end{equation} 
with the transmission of the tomography setup $\eta_{\text{Tomo}}$ = 86.5\% and the quantum efficiencies of the SSPDs $\eta_{\text{SSPD1}}$ = 70(2)\% and $\eta_{\text{SSPD2}}$ = 62(2)\%.\\
Again as a consistency check, we calculate the expected rate with the losses and efficiencies in the telecom setup measured with a laser. For the overall transmission efficiency we get 
\begin{equation}
\eta_{\text{1310,trans}} = \eta_{\text{Fiber}}\cdot\eta_{\text{ExCon}}\cdot\eta_{\text{Stab}} = 17.5(1)\% 
\end{equation}
with the transmission of the fiber between the labs (includes the coupling to the fiber and the fiber-fiber coupling to the converter) $\eta_{\text{Fiber}}$ = 75.8\%, the external conversion efficiency $\eta_{\text{ExCon}}$ = 26.5(2)\% and the duty cycle of the path-length stabilization $\eta_{\text{Stab}}$ = 87.5\%. With a sequence repetition rate of $\gamma_{\text{rep}} \approx$\,61.7\,kHz (the rate is slightly increased as the atomic state analysis is performed less often caused by a lower photon count rate) the expected generated rate is   
\begin{equation}
 \gamma_{\text{1310,gen,theo}} = \gamma_{\text{rep}}\cdot \eta_{\text{Halo}}\cdot \eta_{\text{Fiber}} \cdot\eta_{\text{Mix}} \cdot\eta_{\text{850}} \cdot\eta_{\text{Phot}} \cdot\eta_{\text{1310,trans}} \approx 44\,\text{events/s}
\end{equation} 
which is again consistent with the measured number. A further interesting quantity, which is important for the following SBR calculation, is the ratio of the probabilities to project and detect a photon at 854\,nm and at 1310\,nm:    
\begin{equation}
 \frac{\eta_{1310, tot}}{\eta_{854, tot}} = \frac{\eta_{\text{1310,trans}} \cdot\eta_{\text{Tomo}} \cdot\frac{\eta_{\text{SSPD1}}+\eta_{\text{SSPD2}}}{\text{2}}}{\eta_{\text{Pol,trans}} \cdot\eta_{\text{APD}}} = 0.427(15)
\end{equation} 
Compared to $\eta_{\text{1310,trans}}$ this value is clearly higher caused by the higher quantum efficiencies of the SSPDs. Note that for the measurement including the projection to the Bell state all numbers have to be multiplied by a factor 2/3 as losses were introduced to compensate the different CGC.

%%%%%%%%%%%%%%%%%%%%%%%%%%%%%%
\vspace{3em}
\Large\noindent\textbf{Supplementary Note 5\vspace{0.3em}\\ Signal-to-background ratio}\normalsize\\

Without the frequency converter we achieve a signal-to-background ratio defined as the total number of signal events (S$_{\text{854}}$ = 114,200 in 4,132\,s inside a 300\,ns window) divided by the total number of background events (BG$_{\text{854}}$ = 3,868) of 29.5. The background in this case is solely determined by the APD dark counts, which is clearly visible if we calculate the expected background yielding    
\begin{equation}
	\text{BG}_{\text{854, theo}} = \text{DC}_{\text{APD}}\cdot 300\,\text{ns} \cdot\eta_{\text{Mix}} \cdot\eta_{\text{850}} \cdot\gamma_{\text{rep}}\cdot 4,132\,s = 3,803 
\end{equation} 
which is in good agreement with the measured background. $DC_{\text{APD}}$ denotes the APD dark-count rate of 117.7\,photons/s.\\

A similar result is obtained for the case including the frequency conversion. The number of signal events (S$_{\text{1310}}$ = 193,120 in 7,779\,s inside a 300\,ns window) and background events (BG$_{\text{1310}}$ = 7,953) yields a SBR of 24.3. Here the background has two contributions; a major part are the SSPD dark counts (on average 93.5\%) and a minor part the conversion-induced noise of 11.4\,photons/s  (on average 6.5\%):   
\begin{equation}
	\text{DC}_{\text{1310\_1}} = \text{DC}_{\text{SSPD1}} + \frac{\text{DC}_{\text{Conv}}}{2}\cdot\eta_{\text{SSPD1}} = 58.7 + \frac{11.4}{2} \cdot 0.7 = 62.7(2)\,\text{photons/s} 
\end{equation} 
\begin{equation}
	\text{DC}_{\text{1310\_2}} = \text{DC}_{\text{SSPD2}} + \frac{\text{DC}_{\text{Conv}}}{2}\cdot\eta_{\text{SSPD2}} = 56.4 + \frac{11.4}{2} \cdot 0.62 = 59.9(2)\,\text{photons/s}
\end{equation} 
A similar calculation reveals that these dark counts determine the total background:  
\begin{equation}
	\text{BG}_{\text{1310, theo}} = \left(\text{DC}_{\text{1310\_1}} + \text{DC}_{\text{1310\_2}} \right) \cdot 300\,\text{ns} \cdot\eta_{\text{Mix}} \cdot\eta_{\text{850}} \cdot\gamma_{\text{rep}}\cdot 7,779\,s = 7,935(25) 
\end{equation} 

As a conclusion, inserting the frequency converter lowers the SBR from 29.5 to 24.3. Although we detect only $\approx$ 43\% of the photons at 1310\,nm, the SBR is not reduced by this factor as the background is also reduced by a factor 0.52, compensating the losses partially. Nevertheless, further improvements of the SBR are reasonable to distribute entanglement over distances of several hundreds of kilometers. On the one hand, the conversion-induced noise can be reduced by spectral filters with smaller bandwidths. Commercial fiber Bragg gratings offer bandwidths down to 1\,GHz without a significant decrease in transmission (Ultra narrow-band filter, Advanced optical solutions GmbH), which already increases the SBR by a factor 25 assuming that the noise is equally distributed around the target wavelength. The fundamental limit is the photon bandwidth of 23\,MHz, but in this regime a tradeoff between filter transmission and filter bandwidth has to be accepted. On the detector side, recent progress in the development of NbTiN-based superconducting detectors with milli-Hertz dark count rates \cite{tangA} are very promising to further increase the SBR about two orders of magnitude compared to our results.                  

\vspace{3em}
\Large\noindent\textbf{Supplementary Note 6\vspace{0.3em}\\ State reconstruction}\normalsize\\ 

\begin{figure*}[ht]
	\centering
	\includegraphics[width=0.9\textwidth]{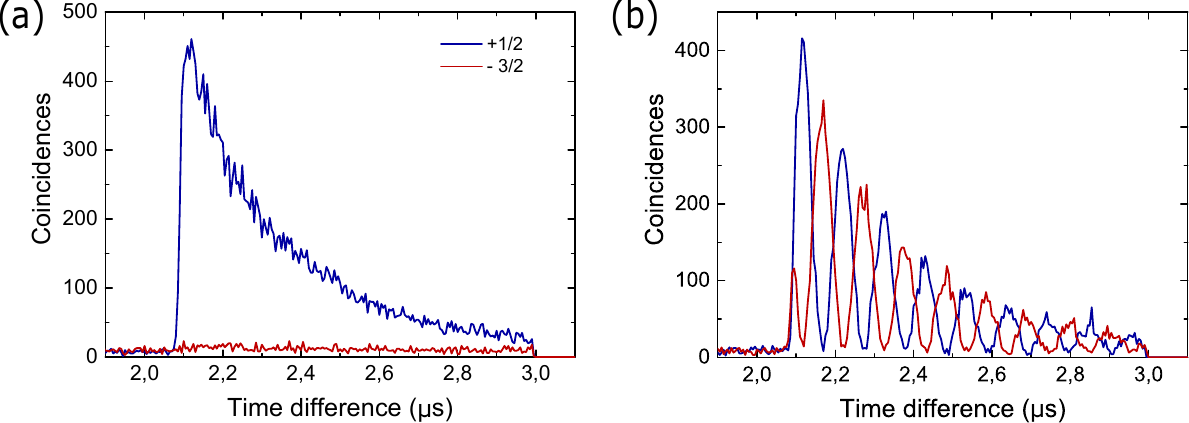}
	\caption{\textbf{Coincidences between 1310\,nm photon detection events and bright events from the atomic state analysis.} (a)~ 1310\,nm photon and ion are projected onto one of the eigenbases, i.e. the photon is projected to $\ket{L}$ and the ion to $\sigma_{z}$. Thus, no $\pi/2$ radio-frequency (RF) pulse resonant to the $\Ket{S_{1/2}, m=-\nicefrac{1}{2}} \rightarrow \Ket{S_{1/2}, m=+\nicefrac{1}{2}}$ transition is applied. (b)~ Projection into one of the superposition bases, i.e. the photon is projected to $\ket{V}$ and the ion to $\sigma_{x}$/$\sigma_{y}$ realized by the RF $\pi/2$-pulse.}
	\label{fig:phot}
\end{figure*}

The state reconstruction is adapted from \cite{kurz1A, kurz2A, kwiat}. We start with the coincidences between photon-detection events and bright events from the atomic state analysis measured in all 36 combinations of projections to the eigenstates of the Pauli operators $\left\{\pm\sigma_{x}, \pm\sigma_{y}, \pm\sigma_{z} \right\}$. Supplementary Figure~\ref{fig:phot}a shows an example for the measurement in the eigenbasis $\sigma_{z}$ where the 1310\,nm photon is projected to $\ket{L}$ and the ion to $\Ket{m=-\nicefrac{3}{2}}$ (red curve) and $\Ket{m=\nicefrac{1}{2}}$ (blue curve), respectively. One measurement in the superposition basis is shown in Supplementary Figure~\ref{fig:phot}b (photon projected to $\ket{V}$). The oscillation is related to the Larmor precession of the phase of the atomic superposition in the $D_{5/2}$-manifold. As the time difference between the start of the excitation pulse at 393\,nm and the atomic state analysis is constant for each sequence repetition, the phase of the atomic superposition depends on the photon's detection time, i.e. the time at which the atom-photon entanglement is established.             

From the 36 measurements we calculate the normalized joint expectation values $S_{i,j} = \braket{\sigma_{i}\otimes\sigma_{j}}$ in the Pauli base $\left\{1, \sigma_{x}, \sigma_{y}, \sigma_{z}\right\}$. This calculation is straightforward for the eigenbases, where we extract solely the number of coincidences (summed up in the 300\,ns window) from the curves shown above. In the case of the projection to the atomic superpositions, the expectation values are determined by the phase and visibility of the Larmor fringes, which we obtain via a fit of the data. With the $S_{i,j}$ we get for the density matrix
\begin{equation}
	\rho = \frac{1}{4} \sum_{i,j=0}^{3} S_{i,j} \sigma_{i}\otimes\sigma_{j}
\end{equation}     
To ensure that the density matrix is physical, a maximum-likelihood estimation \cite{kwiat} is performed afterwards with the linear reconstructed matrix as initial guess. This final density matrix is utitlized to calculate the fidelities and purities stated in the main text and the next section. The error bars of the latter are derived via a finite difference-quotient method assuming a Poissonian distribution of the detection events. 

\vspace{3em}
\Large\noindent\textbf{Supplementary Note 7\vspace{0.3em}\\ Complete list of fidelities and purities}\normalsize\\ 

As mentioned in the main text, background subtraction was applied to the raw data to demonstrate the functionality of our method. To perform this background subtraction, we reconstructed the density matrix as described in In Supplementary Note 6 from a modified raw data set. At first, we determine for each basis combination the average number of dark coincidences per time-bin. In Supplementary Figure~\ref{fig:phot}a, for instance, this is done by integrating the number of coincidences outside the photon wavepacket (from 0\textmu s to 2\textmu s) and dividing it by the number of time-bins. This approach is justified in our experiment as we generate triggered single photons within a specified time window; thus, contrary to a photon-pair source, e.g. based on parametric conversion with random emission times, coincidences outside the photon wavepacket originate solely from detector dark-counts. Now, we assume a Poissonian distribution for the dark coincidences (this is feasible as conversion-induced noise as well as detector dark counts occur independently) with the average number of dark coincidences as expectation value. With the distribution of the dark coincidences and the total number of coincidences, we calculate the distribution of the true signal coincidences, yielding expectation value and variance of the latter. From this point, we proceed as described in Supplementary Note 6. This statistical approach is necessary on the one hand to get a proper error estimation and on the other hand to avoid negative values for the number of coincidences in some time-bins.\\ 

The following tables summarize again all fidelities and purities with and without subtraction of the background (denoted in the tables as "Total BG subt."' and "W/o BG subt."', respectively). In the measurements with the frequency converter, 93.5\% of the background originates from the detectors and 6.5\% from conversion-induced noise. Thus it is of interest to quantify solely the converter's influence on the state by subtracting only 93.5\% of the background (in this case we multiply the average number of dark coincidences per bin with 0.935). This is denoted in the tables as "Detector DC subt."

\begin{table}[ht]
	\centering
	\caption{The fidelities denoting the overlap with the expected state}
	\vspace{0.5em}
	\begin{tabular}{cccc}
		\toprule
		Measurement & Total BG subt. & Detector DC subt. & W/o BG subt. \\ \midrule
		W/o conversion  & 98.3 $\pm$ 0.3 \% & -- & 95.9 $\pm$ 0.3 \% \\
		With conversion  & 97.7 $\pm$ 0.2 \% & 97.3 $\pm$ 0.2 \% & 94.8 $\pm$ 0.2 \% \\ \bottomrule 
	\end{tabular}
\end{table}

\begin{table}[ht]
	\centering
	\caption{The fidelities denoting the overlap with the Bell state}
	\vspace{0.5em}
	\begin{tabular}{cccc}
		\toprule
		Measurement & Total BG subt. & Detector DC subt. & W/o BG subt. \\ \midrule
		W/o conversion  & 95.5 $\pm$ 0.3 \% & -- & 93.3 $\pm$ 0.3 \%  \\
		With conversion  & 94.8 $\pm$ 0.2 \% & 94.5 $\pm$ 0.2 \% & 92.2 $\pm$ 0.2 \%  \\
		With conv. \& state carving   & 98.2 $\pm$ 0.2 \% & 97.7 $\pm$ 0.2 \%  & 93.4 $\pm$ 0.2 \%\\ \bottomrule
	\end{tabular}
\end{table}

\begin{table}[ht]
	\centering
	\caption{The purities of all measurements}
	\vspace{0.5em}
	\begin{tabular}{cccc}
		\toprule
		Measurement & Total BG subt. & Detector DC subt. & W/o BG subt. \\ \midrule
		W/o conversion  & 96.7 $\pm$ 1.6 \% & -- & 92.1 $\pm$ 1.6 \%  \\  
		With conversion  & 95.8 $\pm$ 1.3 \% & 95.1 $\pm$ 1.3 \% & 90.3 $\pm$ 1.2 \%  \\
		With conv. \& state carving   & 96.7 $\pm$ 1.4 \% & 95.8 $\pm$ 1.4 \%  & 87.8 $\pm$ 1.3 \%\\ \bottomrule
	\end{tabular}
\end{table}

\FloatBarrier

%% Dokument ENDE %%%%%%%%%%%%%%%%%%%%%%%%%%%%%%%%%%%%%%%%%%%%%%%%%%%%%%%%%%
\end{document}